\documentclass[12pt]{article}
\usepackage{fullpage,authblk,cite}
\usepackage{amsmath,amsthm,graphicx,amssymb,verbatim,listings}
\usepackage{wasysym,color,enumitem,mathcomp}
\usepackage[T1]{fontenc}     % needed for the guillemets
\usepackage{lmodern, tipa}   % needed to fix suboptimal font rendering
\usepackage[ pdftex, plainpages = false, pdfpagelabels,
                 pdfpagelayout = useoutlines,
                 bookmarks,
                 bookmarksopen = true,
                 bookmarksnumbered = true,
                 breaklinks = true,
                 linktocpage=all,
                 pagebackref=false,
                 colorlinks = true,
                 linkcolor = Blue,
                 urlcolor  = blue,
                 citecolor = BrickRed,
                 anchorcolor = green,
                 hyperindex = true,
                 hyperfigures
                 ]{hyperref}
\usepackage[usenames, dvipsnames]{xcolor}
\usepackage{xifthen} % provides \isempty test

\newcommand{\arxiv}[2][]{\ifthenelse{\isempty{#1}}{\href{http://arxiv.org/abs/#2}{{\tt arXiv:\allowbreak{}#2}}} {\href{http://arxiv.org/abs/#2}{{\tt arXiv:\allowbreak{}#2 [#1]}}}}

\newcommand{\booktitle}{\textsl}
\newcommand{\hrefdoi}[2]{\href{https://dx.doi.org/#1}{#2}}

\begin{document}
\title{A Critical, But Hopefully Cordial, QBist Reply to Ballentine}
\author[$\dag$]{Blake C.\ Stacey}
\affil[$\dag$]{Physics Department, University of
    Massachusetts Boston\protect\\ 100 Morrissey Boulevard, Boston MA 02125, USA}

\date{\small\today}

\maketitle

\begin{abstract}
L.\ E.\ Ballentine's remarks in \booktitle{Physics Today} about the
QBist interpretation of quantum mechanics are generally wide of the
mark.
\end{abstract}

In 2019, DiVincenzo and Fuchs wrote an article for \booktitle{Physics
  Today} about the storied history of quantum-foundations papers that
have been published in \booktitle{Reviews of Modern
  Physics}~\cite{DiVincenzo:2019}. Ballentine's 1970 article on his
ensemble interpretation of quantum mechanics~\cite{Ballentine:1970}
was inadvertently omitted, not out of a lack of interest in ensemble
interpretations but due to an oversight, failing to recall at the time
that that paper had appeared in \booktitle{RMP}
specifically~\cite{DiVincenzo:2020}. Ensemblist views of course occupy
an important place in the phylogeny of quantum interpretations; they
comport naturally with a very physicist-y intuition for how the
concept of probability ought to be cleared up. Ballentine's letter to
\booktitle{Physics Today} pointing out the
omission~\cite{Ballentine:2020} goes on to raise some other points
that deserve addressing.

Much of Ballentine's letter is a complaint about the QBist
interpretation of quantum mechanics. For canonical references on
QBism, see~\cite{Fuchs:2013, Fuchs:2013b, Fuchs:2016,
  DeBrota:2020}. The responses that follow are largely more focused
versions of explanations made elsewhere, such as in these papers and
the supplemental FAQ~\cite{DeBrota:2018}.

Ballentine writes,
\begin{quote}
  In QM, a measurement of an observable should yield an eigenvalue of
  the observable.
\end{quote}
This is, from our perspective, rather an old-fashioned way of thinking
about quantum mechanics. To a QBist, quantum theory is an
empirically-molded addition to decision theory --- a manual for an
agent to better manage her own expectations about the consequences of
her actions, in light of the character of the physical
world. Therefore, the careful and expensive actions that Alice might
carry out in a laboratory are on a continuum with those actions that
she contemplates doing in her daily life. The old von Neumann idea
that a measurement must correspond to a self-adjoint operator is just
too limited; it stands in the way of appreciating the lessons that
quantum theory has to offer. The experiments that a
quantum-information text might call ``generalized measurements'' are,
to a QBist, the natural notion of \emph{measurement} to begin
with~\cite{Fuchs:2017}.

\begin{quote}
  The usual role of an interpretation of QM is to begin with the
  established mathematical formalism and provide an intuitively
  comprehensible idea of the physical process that the math describes.
\end{quote}
An interpretation of QM provides a narrative about the mathematics,
tying it somehow to physical reality, but whether that narrative is
``intuitively comprehensible'' is not so
clear~\cite{Schaffer:2021}. What is ``intuitively comprehensible''
about causation backwards in time, or an infinity of branching worlds,
or anything that Bohr said about complementarity?

\begin{quote}
  QBism begins with the assumption that all kinds of probability can
  be regarded as subjective Bayesian probabilities. That assumption
  can be maintained only by ignoring the literature on interpretations
  of probability, from which it is clear that several different
  kinds---or interpretations---of probability exist. DiVincenzo and Fuchs
  may have ignored the classic philosophical writings on the subject
  because they were written by philosophers for philosophers and so do
  not address the needs of physicists.
\end{quote}
I can't speak for DiVincenzo, but Fuchs has been reading the
philosophical literature on probability for quite some time. See, for
example, the correspondence dating to the 1990s collected in his
\booktitle{Coming of Age with Quantum
  Information}~\cite{Fuchs:2010}. I, too, have spent more time in that
part of the library than is typical or perhaps even healthy for a
physicist~\cite{Stacey:2016, Stacey:2019b}.

\begin{quote}
  In general, QM states do not determine the results of a measurement,
  only the probabilities of the possible results. That a state's
  influence on the results is not deterministic suggests strongly that
  the quantum probabilities given by the Born rule should be
  interpreted as propensities.
\end{quote}
This appears to stem from a view in which quantum states are more
fundamental than probabilities, i.e., that the latter must be derived
from the former. QBism does not subscribe to this
view~\cite{Fuchs:2017}. Instead, quantum states and probabilities have
the same status. A quantum state, even a pure state, can be
represented equivalently as a probability distribution over the
outcomes of a reference measurement. Or, more provocatively: a quantum
state just \emph{is} a probability distribution, written in a manner
that is convenient for textbook physics calculations but not the best
for conceptual questions. Rather than being a formula that computes
probabilities from a ghostly underlying quantity, the Born rule
expresses how probabilities for different experiments mesh together in
a nonclassical way~\cite{Fuchs:2017, DeBrota:2021, Stacey:2022}.

(One can draw a pedagogical analogy here with relativity. Minkowski
spacetime is a conceptual contrivance that allows a physicist to tie
together expectations about all sorts of experiments involving all
kinds of rods and clocks~\cite{Mermin:2013}. It provides a means to do
this consistently while rejecting the notion of absolute simultaneity,
despite this notion seeming fundamental to everyday intution and
pre-relativistic science. Quantum mechanics makes a corresponding but
more radical move. The quantum formalism lets a physicist interweave
expectations for different experiments in a self-consistent manner
that rejects the idea of underlying hidden variables, despite that
idea seeming so essential in the years before the revolution.)

Ballentine continues,
\begin{quote}
  They [quantum probabilities] refer objectively to the physical
  system and its environment, not to any agent's knowledge, so they
  are not naturally interpreted as subjective Bayesian probabilities.
\end{quote}
This asserts that quantum probabilities cannot be subjective because
they just aren't subjective. The desired conclusion is simply assumed.

\begin{quote}
  E.\ T.\ Jaynes was a well-known supporter of the Bayesian theory of
  probability. In 1989 he repeated Bell's derivation of [his] inequality but
  carefully treated all instances of probability as Bayesian. He found
  that the derivation could not be completed without invoking an extra
  assumption that was not justifiable in the Bayesian theory.
\end{quote}
Within the niche of physicists who care about attempts to knock down
Bell's theorem, those in the sub-niche who have found Jaynes'
take~\cite{Jaynes:1989} worth evaluating have concluded that in this
aspect, he lost the plot~\cite{Gill:2002}. Nothing prevents one from
going through a standard derivation, like Nielsen and Chuang's
presentation of the Bell--CHSH inequality~\cite{Nielsen:2010} or
Mermin's improvement upon the GHZ scenario~\cite{Mermin:1990,
  Mermin:1993}, while treating all probabilities in a personalist
Bayesian manner~\cite{Stacey:2021}. One posits the existence of an
intrinsic hidden variable $\lambda$; an agent's beliefs about what
values the hidden variable might take are encapsulated by a gambling
commitment, a probability distribution $p(\lambda)$. From this, one
deduces constraints on other probabilities and expectations. Then, one
shows that quantum mechanics can violate those constraints. The
conclusion is not to reject Bayesianism, but to reject $\lambda$-ism.

(If one wishes to emulate the later Bell and start instead with a
factorization condition~\cite{Zukowski:2014}, the moral is the same:
Bolting an extra assumption onto probability theory implies
constraints that quantum theory is not obligated to respect.)

QBism does not adopt the Jaynesian version of Bayesian
probability. Instead, QBists occupy the Ramsey/de Finetti portion of
the spectrum, employing personalist rather than objective
Bayesianism~\cite{Stacey:2019}. Ballentine's letter is not the most
overt instance of someone making up an interpretation to get angry
at~\cite{Fuchs:2020}, but it does have a drop of that flavor,
criticizing the author's own extrapolation of the word ``Bayesian''
more than anything that the QBists have actually written.

One passage in Ballentine's letter is a bit of a puzzle:
\begin{quote}
They assert that ``physicists and philosophers are still debating what
a `measurement' really means.'' What is important for [quantum
  foundations] is not the meaning of the word but an understanding of
the physical process. The authors do not cite any of the published
papers that provide such an understanding.
\end{quote}
The assertion in question is a simple statement of fact. Plenty of
people have become physicists, even good physicists, without caring
about \emph{meaning}, sometimes dismissing it as pub fare, or a
dorm-room enthusiasm that one grows out of. But practically by
definition, quantum foundations is an area where one has to care. The
overlap of physics and philosophy will contain at least a little
philosophy. For QBists, the problem with the entire genre of papers
about the ``physical process'' of measurement or ``measurement
models'' is that they only defer and obfuscate the central questions,
instead of resolving them. One can couple system $A$ to system $B$,
and system $B$ to system $C$, and on and on, but this merely
procrastinates. It is like trying to answer the question of what
probability \emph{is} by adding more dice.

Ballentine refers to an earlier paper of his ``on the foundations of
probability theory, written from the point of view of a quantum
physicist''~\cite{Ballentine:2016}. The authors he cites are Fine,
himself, James, Popper, Humphreys, Cox, R\'enyi, McCurdy, Gillies,
Greenberger, Hentschel, Weinert, Bell, Wigner, Araki, Yanase, Ohira,
Pearle, Norsen, Tumulka, Pearl and Penrose.  Notably absent from the
bibliography: Ramsey, de Finetti, Hosiasson, Savage, Hesse, van
Fraassen, Goldstein, Shafer, Diaconis, Skyrms\ldots\ Ballentine brings
up an example of a bent coin and says that, to a Bayesian, a
particular quantity would have to be ``a degree of belief about a
degree of belief''. Then he writes, ``Perhaps such a recursive belief
can be made sense of''; due to the paucity of references, it is not
clear why the lengthy tradition of studying such recursion in the
personalist framework would be unsatisfactory. (A paper by Fuchs and
Schack~\cite{Fuchs:2011} and the textbook by Diaconis and
Skyrms~\cite{Diaconis:2018} provide two entry points into this
literature. It is also touched upon in Home and Whitaker's critical
review of ensemble interpretations~\cite{Home:1992}.) Moreover,
Ballentine seems to conflate the matter of how people actually act
with the question of the \emph{normative} standards which they should
\emph{strive} to meet~\cite{Fuchs:2017}.

In \S 5.4 of his 2016 paper, Ballentine presents an argument for why
quantum probabilities cannot be subjective. His scenario involves two
physicists, Alice and Bob, who ``have different information'' about a
photon but write the same density matrix. Ballentine's argument fails
to go through, because the quantities that Alice and Bob write down
differently (equivalent to a choice of POVM elements) are \emph{also}
personalist gambling commitments to a QBist. Ballentine's
thought-experiment boils down to a pair of agents with differing
\emph{overall} beliefs changing those beliefs in different ways, which
is not surprising at all.

(The description of the thought-experiment cheats a little bit by
introducing a third party who knows what the true, physical, objective
state of the photon is. In other words, it is yet another
thought-experiment that asks, ``How can quantum states be subjective
if we assume that they are objective?'' It is hard to tell how such
arguments can end up being very illuminating~\cite{Stacey:2023}.)

In summary, Ballentine's critiques fail to connect and instead retread
old ground. This is a missed opportunity, since QBism is an ongoing
project, and new criticism is part of what moves a project forward.


\begin{thebibliography}{999}

\bibitem{DiVincenzo:2019} D.\ P.\ DiVincenzo and C.\ A.\ Fuchs,
  ``\hrefdoi{10.1063/PT.3.4141}{Quantum foundations},''
  \booktitle{Physics Today} \textbf{72}, 2 (2019), 50--51.

\bibitem{Ballentine:1970} L.\ E.\ Ballentine,
  ``\hrefdoi{10.1103/RevModPhys.42.358}{The Statistical Interpretation
  of Quantum Mechanics},'' \booktitle{Reviews of Modern Physics}
  \textbf{42} (1970), 358.

\bibitem{DiVincenzo:2020} D.\ P.\ DiVincenzo and C.\ A.\ Fuchs,
  ``\hrefdoi{10.1063/PT.3.4141}{Reviews of quantum foundations},''
  \booktitle{Physics Today} \textbf{73}, 6 (2020), 12.

\bibitem{Ballentine:2020} L.\ E.\ Ballentine,
  ``\hrefdoi{10.1063/PT.3.4488}{Reviews of quantum foundations},''
  \booktitle{Physics Today} \textbf{73}, 6 (2020), 11--12.
  
\bibitem{Fuchs:2013} C.\ A.\ Fuchs, N.\ D.\ Mermin and R.\ Schack,
  ``\hrefdoi{10.1119/1.4874855}{An introduction to QBism with an
  application to the locality of quantum mechanics},''
  \booktitle{American Journal of Physics} \textbf{82} (2014), 749--54,
  \arxiv{1311.5253}.

\bibitem{Fuchs:2013b} C.\ A.\ Fuchs and R.\ Schack,
  ``\hrefdoi{10.1103/RevModPhys.85.1693}{Quantum-Bayesian
  coherence},'' \booktitle{Reviews of Modern Physics} \textbf{85}
  (2013), 1693--1715, \arxiv{1301.3274}.

\bibitem{Fuchs:2016} C.\ A.\ Fuchs and B.\ C.\ Stacey,
  ``QBism:\ Quantum theory as a hero's handbook.''
  \booktitle{Proceedings of the International School of Physics
    ``Enrico Fermi,'' Course 197 -- Foundations of Quantum Physics},
  edited by E.\ M.\ Rasel, W.\ P.\ Schleich, and S.\ W\"olk (IOS
  Press, 2019). \arxiv{1612.07308}.
  
\bibitem{DeBrota:2020} J.\ B.\ DeBrota, C.\ A.\ Fuchs and R.\ Schack,
  ``\hrefdoi{10.1007/s10701-020-00369-x}{Respecting one's fellow:
  QBism's analysis of Wigner's friend},'' \booktitle{Foundations of
  Physics} \textbf{50} (2020), 1859--74, \arxiv{2008.03572}.

\bibitem{DeBrota:2018} J.\ B.\ DeBrota and B.\ C.\ Stacey, ``FAQBism,''
  \arxiv{1810.13401} (2018).

\bibitem{Fuchs:2017} C.\ A.\ Fuchs, ``Notwithstanding Bohr, the
  Reasons for QBism,'' \booktitle{Mind and Matter} \textbf{15} (2017),
  245--300, \arxiv{1705.03483}.

\bibitem{Schaffer:2021} K.\ Schaffer and G.\ Barreto Lemos,
  ``\hrefdoi{10.1007/s10699-019-09608-5}{Obliterating Thingness: An
  Introduction to the `What' and the `So What' of Quantum Physics},''
  \booktitle{Foundations of Science} \textbf{26} (2021), 7--26,
  \arxiv{1908.07936}.
  
\bibitem{Fuchs:2010} C.\ A.\ Fuchs, \booktitle{Coming of Age with
  Quantum Information} (Cambridge University Press, 2011).

\bibitem{Stacey:2016} B.\ C.\ Stacey, ``Von Neumann Was Not a Quantum
  Bayesian,'' \booktitle{Philosophical Transactions of the Royal
    Society A} \textbf{374} (2016), 20150235, \arxiv{1412.2409}.

\bibitem{Stacey:2019b} B.\ C.\ Stacey,
  ``\hrefdoi{10.1387/theoria.20465}{Book Review:\ \booktitle{What Is
    Quantum Information?}},'' \booktitle{Theoria} \textbf{34} (2019),
  153--55.

\bibitem{DeBrota:2021} J.\ B.\ DeBrota, C.\ A.\ Fuchs, J.\ L.\ Pienaar and
  B.\ C.\ Stacey, ``\hrefdoi{10.1103/PhysRevA.104.022207}{Born's rule
    as a quantum extension of Bayesian coherence},''
  \booktitle{Physical Review A} \textbf{104} (2021),
  022207. \arxiv{2012.14397}.

\bibitem{Stacey:2022} B.\ C.\ Stacey, ``The Status of the Bayes Rule
  in QBism,'' \arxiv{2210.10757} (2022).

\bibitem{Mermin:2013} N.\ D.\ Mermin, ``QBism as CBism:\ Solving the
  Problem of `the Now','' \arxiv{1312.7825} (2013).
  
\bibitem{Jaynes:1989} E.\ T.\ Jaynes, ``Clearing up mysteries---the
  original goal.'' In \booktitle{Maximum Entropy and Bayesian
    Methods,} edited by J.\ Skilling (Kluwer, 1989).

\bibitem{Gill:2002} R.\ D.\ Gill, ``Time, Finite Statistics, and
  Bell's Fifth Position.'' In \booktitle{Proceedings of the Conference
    Foundations of Probability and Physics -- 2 : V\"axj\"o (Soland),
    Sweden, June 2--7, 2002.} (V\"axj\"o University Press, 2002.)
  \arxiv{quant-ph/0301059}.

\bibitem{Nielsen:2010} M.\ A.\ Nielsen and I.\ L.\ Chuang,
  \booktitle{Quantum Computation and Quantum Information} (Cambridge
  University Press, 2010), pp.\ 111--17.
  
\bibitem{Mermin:1990} N.\ D.\ Mermin,
  ``\hrefdoi{10.1119/1.16503}{Quantum mysteries revisited},''
  \booktitle{American Journal of Physics} \textbf{58} (1990), 731--34.
  
\bibitem{Mermin:1993} N.\ D.\ Mermin,
  ``\hrefdoi{10.1103/RevModPhys.65.803}{Hidden variables and the two
  theorems of John Bell},'' \booktitle{Reviews of Modern Physics}
  \textbf{65} (1993), 803--15, \arxiv{1802.10119}.

\bibitem{Stacey:2021} B.\ C.\ Stacey,
  \booktitle{\hrefdoi{10.1007/978-3-030-76104-2}{A First Course in the
      Sporadic SICs}} (Springer, 2021), pp.\ 40--41.

\bibitem{Zukowski:2014} M. \.{Z}ukowski and \v{C}.\ Brukner,
  ``\hrefdoi{10.1088/1751-8113/47/42/424009}{Quantum non-locality---it
  ain't necessarily so\ldots},'' \booktitle{Journal of Physics A}
  \textbf{47} (2014), 424009, \arxiv{1501.04618}.
  
\bibitem{Stacey:2019} B.\ C.\ Stacey, ``Ideas Abandoned en Route to
  QBism,'' \arxiv{1911.07386} (2019).

\bibitem{Fuchs:2020} C.\ A.\ Fuchs and B.\ C.\ Stacey, ``QBians Do Not
  Exist,'' \arxiv{2012.14375} (2020).

\bibitem{Ballentine:2016} L.\ E.\ Ballentine,
  ``\hrefdoi{10.1007/s10701-016-9991-0}{Propensity, Probability, and
  Quantum Theory},'' \booktitle{Foundations of Physics} \textbf{46}
  (2016), 973--1005.

\bibitem{Fuchs:2011} C.\ A.\ Fuchs and R.\ Schack,
  ``\hrefdoi{10.1007/978-3-642-21329-8_15}{Bayesian conditioning, the
  reflection principle, and quantum decoherence}.'' In
  \booktitle{Probability in Physics}, edited by Y.\ Ben-Menahem and
  M.\ Hemmo (Springer, 2012). \arxiv{1103.5950}.

\bibitem{Diaconis:2018} P.\ Diaconis and B.\ Skyrms, \booktitle{Ten
  Great Ideas about Chance} (Princeton University Press, 2018).

\bibitem{Home:1992} D.\ Home and M.\ Whitaker,
  ``\hrefdoi{10.1016/0370-1573(92)90088-H}{Ensemble interpretations of
  quantum mechanics.\ A modern perspective},'' \booktitle{Physics
  Reports} \textbf{210} (1992), 223--317.
  
\bibitem{Stacey:2023} B.\ C.\ Stacey, ``Whose Probabilities? About
  What? A Reply to Khrennikov,'' \arxiv{2302.09475} (2023).
  
\end{thebibliography}
\end{document}